\documentclass[prb,aps,showpacs,amsmath,amssymb,superscriptaddress,affilletter,twocolumn]{revtex4-1}

\usepackage[usenames,dvipsnames]{color}
\usepackage{graphicx}
\usepackage{tabularx}
\usepackage[T1]{fontenc}
\usepackage[cp1250]{inputenc}
\usepackage{dcolumn}
\usepackage{amsmath,amsfonts,amssymb}
\usepackage{multirow}
\usepackage[rightcaption]{sidecap}
\usepackage{float}
\newcommand{\spl}{\sigma^+}
\newcommand{\smn}{\sigma^-}

\newcommand{\CdMnTe}{Cd$_{1-x}$Mn$_x$Te}
\newcommand{\CdMnSe}{Cd$_{1-x}$Mn$_x$Se}
\newcommand{\CdZnTe}{Cd$_{1-y}$Zn$_y$Te}
\newcommand{\ZnMnTe}{Zn$_{1-x}$Mn$_x$Te}

\newcommand{\mean}[1]{\langle{#1}\rangle{}}
\newcommand{\ket}[1]{\left|#1\right>}
\newcommand{\bra}[1]{\left< #1 \right|}
\newcommand{\beq}{\begin{equation}}
\newcommand{\eeq}{\end{equation}}

\begin{document}

\title{Influence of exciton spin relaxation on the photoluminescence spectra \\ of semimagnetic quantum dots}

\author{Ł.~Kłopotowski}
\email[Corresponding author: ]{lukasz.klopotowski@ifpan.edu.pl}
\affiliation{Institute of Physics, Polish Academy of Sciences, Al. Lotników 32/46, 02-668 Warsaw, Poland}

\author{Ł.~Cywiński}       \affiliation{Institute of Physics, Polish Academy of Sciences, Al. Lotników 32/46, 02-668 Warsaw, Poland}

\author{M.~Szymura}         \affiliation{Institute of Physics, Polish Academy of Sciences, Al. Lotników 32/46, 02-668 Warsaw, Poland}

\author{V. Voliotis}
\affiliation{Institut des NanoSciences de Paris, Universit{\'e} Pierre et Marie Curie, CNRS, 4 place Jussieu, 75252 Paris, France}

\author{R. Grousson}
\affiliation{Institut des NanoSciences de Paris, Universit{\'e} Pierre et Marie Curie, CNRS, 4 place Jussieu, 75252 Paris, France}

\author{P.~Wojnar}          \affiliation{Institute of Physics, Polish Academy of Sciences, Al. Lotników 32/46, 02-668 Warsaw, Poland}

\author{K.~Fronc}           \affiliation{Institute of Physics, Polish Academy of Sciences, Al. Lotników 32/46, 02-668 Warsaw, Poland}

\author{T. Kazimierczuk}
\affiliation{Faculty of Physics, University of Warsaw, ul. Hoża 69, 00-681 Warsaw, Poland}

\author{A. Golnik}
\affiliation{Faculty of Physics, University of Warsaw, ul. Hoża 69, 00-681 Warsaw, Poland}

\author{G.~Karczewski}      \affiliation{Institute of Physics, Polish Academy of Sciences, Al. Lotników 32/46, 02-668 Warsaw, Poland}

\author{T.~Wojtowicz}       \affiliation{Institute of Physics, Polish Academy of Sciences, Al. Lotników 32/46, 02-668 Warsaw, Poland}

\date{\today}

\begin{abstract}

We present a comprehensive experimental and theoretical studies of  photoluminescence of single \CdMnTe\ quantum dots with Mn content $x$ ranging from 0.01 to 0.2. We distinguish three stages of the equilibration of the exciton-Mn ion spin system and show that the intermediate stage, in which the exciton spin is relaxed, while the total equilibrium is not attained, gives rise to a specific asymmetric shape of the photoluminescence spectrum. From an excellent agreement between the measured and calculated spectra we are able to evaluate the exciton localization volume, number of paramagnetic Mn ions, and their temperature for each particular dot. We discuss the values of these parameters and compare them with results of other experiments. Furthermore, we analyze the dependence of average Zeeman shifts and transition linewidths on the Mn content and point out specific processes, which control these values at particular Mn concentrations.

\end{abstract}


\maketitle

\section{Introduction}

Zero dimensional density of states for carriers confined in self-assembled quantum dots (QDs), make these nanostructures unique in many aspects. Most importantly from the point of view of applications, they offer a combination of the zero-dimensional, atomic-like optical properties with easy integration into the existing semiconductor technology.\cite{mic09} As a result, many proposals emerge on employing quantum dots as sources of classical\cite{ara82} and non-classical light,\cite{mic00} quantum bits (qubits) and quantum gates,\cite{bar95,los98} systems for storage of classical\cite{kro04} and quantum\cite{han07,liu10} information, and ultra-precise sensors of local electric fields.\cite{hou12}

Semimagnetic QDs possess an additional degree of freedom i.e., the magnetic moment associated with the magnetic impurities incorporated into the dot material. Therefore, these dots offer a unique opportunity to combine the zero-dimensional electronic properties with the paramagnetism of the semimagnetic semiconductor. Harnessing the magnetism in these nanostructures can lead to novel nanoscale devices. In particular, electrical control of the QD charge state is expected to result in switching of the QD magnetization on and off.\cite{fer04,gov05,abo07} Optical control of the magnetization\cite{mac04} and its long relaxation time\cite{gur08} can provide a path toward employing semimagnetic dots in memory storage devices. Also, a pair of coupled nonmagnetic and semimagnetic dots was proposed as a qubit gate, with electric field tunability.\cite{lya12}

By far, the most studied semimagnetic QDs are made of II - VI semiconductors, since the low solubility of Mn in III - V compounds hinders introduction of more than one Mn ion into a dot.\cite{kud07} Among the studied materials, the most attention has been devoted to \CdMnTe\ and \CdMnSe. In these dots, the {\em s,p-d} exchange interaction between the charge carriers and localized magnetic ions underlies the optical properties analogously to higher dimension systems.\cite{gaj10} In particular, giant Zeeman splittings of the excitonic sublevels are observed, with a magnitude that is directly proportional to the dot magnetization as evidenced by quantitative agreement between the exciton recombination energies and a modified Brillouin function.\cite{mak00,bac02,hun04} Another striking spectroscopic feature of these dots is a substantial broadening of the photoluminescence (PL) transition.\cite{mak00,bac02} The linewidths at zero magnetic field are on the order of a few meV, whereas for nonmagnetic CdTe and InAs QDs, the transitions are one and two orders of magnitude narrower, respectively.\cite{bes01,gam96} The broadening is another consequence of the exciton-Mn exchange interaction: thermal fluctuations of the Mn spins introduce a temporal magnetic field, fluctuating on the timescale of hundreds of nanoseconds, which in turn influences the transition linewidth in a time averaged spectrum. Remarkably, with increasing magnetic field, the PL transitions become more narrow, as the fluctuations of the Mn spins are suppressed when their Zeeman splitting becomes larger then the thermal energy. Thus, at high magnetic fields, the transition linewidth tends to the one observed for a Mn-undoped, CdTe or CdSe dot.\cite{mak00,bac02,woj07}

Most of the studies on semimagnetic dots involved an assumption of thermal equilibrium between the Mn spin system and the exciton. In such a case, the exchange energy is minimized by spontaneous formation of a ferromagnetic order in the paramagnetic Mn spin system.\cite{mak00,bac02,seu02,bea09,sel10,klo11pol,pie12} This spontaneous magnetization forms within the exciton wave function \cite{kav99} and is hence labeled as excitonic magnetic polaron (EMP). Formation of the EMP is therefore synonymous with reaching the equilibrium in the exciton-Mn spin system. The lineshapes become Gaussian with the full width at half maximum (FWHM) directly monitoring the statistical spin fluctuations in the exchange field of the exciton.\cite{bac02}
However, achieving the equilibrium requires the characteristic timescale of collective Mn magnetization dynamics, closely related to Mn spin-spin relaxation time,\cite{die95} to be  shorter than the exciton recombination time $\tau_r$. In \CdMnTe\ QDs, the latter time is on the order of 300 ps.\cite{klo11pol} Thus, in order to achieve the equilibrium, the Mn magnetization dynamics has to occur on a timescale of 1-100 ps, much shorter than those driven by spin-lattice interactions, which occur on the $\mu$s/ms timescales.\cite{die95,yak96} The spin-spin mechanism becomes efficient only if the average distance between Mn spins is small enough. Consequently, EMP formation was unambiguously demonstrated only for dots with a Mn content $x \geq 0.1$. For dots with a more diluted concentration of Mn spins, EMP starts to form upon photoexcitation, but exciton recombination interrupts the establishment of equilibrium.\cite{klo11pol}

Another important question concerns the possibility of an optical orientation of the EMP. \cite{mac04,gur08} This could only be possible if the exciton spin relaxation time were longer than the EMP formation time $\tau_f$. We have shown that for relatively large self-assembled \CdMnTe\ QDs the opposite is true.\cite{klo11pol} We have established the EMP formation scenario in which the exciton adjusts its spin to the direction of a momentary, initial magnetization fluctuation in the dot and then, provided that $\tau_r > \tau_f$ the EMP is formed. Therefore, we distinguish three stages in reaching the equilibrium. Initially, shortly after the photocreation of the exciton, the system is completely unrelaxed with the exciton spin being either random if the orientation was lost during the capturing by the QD potential, or reflecting the photon angular momentum if otherwise. In either case, the exciton spin has no preferential orientation with respect to the magnetization. In the second stage, the exciton relaxes its spin, which is synonymous with thermal population of the exciton spin states in the exchange field of the Mn spins. The  Mn magnetization is then increasing due to Mn spin-spin interactions until it reaches its equilibrium value for a given exchange field provided by a spin-polarized exciton. At the end of this third stage the full  equilibrium is reached and the EMP is formed.

In this paper, we aim to establish a quantitative description of the optical properties of semimagnetic QDs containing many Mn ions and demonstrate the influence of the three relaxation stages on the shape of the PL spectrum. In particular, we show that the relaxation of the exciton spin has a nontrivial effect on the shape of the PL line and leads to its narrowing by a factor as high as 2 and development of asymmetrical shape in a certain range of temperatures and magnetic fields. We start with presenting the theoretical model, which allows to compute the PL spectra particular for the different relaxation stages. Among model parameters are the number of paramagnetic Mn ions interacting with the exciton and the exciton localization volume, related to the QD volume. Thus, the by fitting the model to the measured PL spectra, we gain access to vital structural parameters describing these semimagnetic dots. We start presenting the experimental results with showing how the PL spectrum evolves in time reaching subsequent stages and confirming our predictions. We then demonstrate the change of the line shape as the magnetic field and temperature is varied. We directly compare the measured spectra with the calculated ones and find an excellent agreement, which allows us to estimate the structural parameters mentioned above. Furthermore, we analyze Zeeman shifts and PL line widths for a large set of QDs from samples with the Mn content varied between 0.01 and 0.2 and discuss them in the framework of the presented model.

\section{Theoretical description}
\label{teos}

\subsection{Distribution of Mn spins in an external field}  \label{sec:Mn_distribution}
In this work, we are interested in statistical properties of a finite number of Mn spins ($N_{\text{Mn}}$) confined in the QD volume.
For a given Mn concentration $x$, $N_{\text{Mn}}$ experiences statistical fluctuations with a standard deviation $\delta N_{\text{Mn}} \! \propto \! \sqrt{N_{\text{Mn}}}$, making it more meaningful to use $N_{\text{Mn}}$ instead of $x$ as an independent variable. We study QDs with $N_{\text{Mn}} \geq 20$ and at moderate magnetic fields, where the magnetization is far from saturation. This allows us to treat the total Mn spin $\mathbf{S} =  \sum^{N_{\text{Mn}}}_i \mathbf{S}_i$ as a classical variable and assume that the probability distribution of $S^{z}$ is Gaussian:
\begin{eqnarray}
P(S^{z}) = \frac{1}{\sqrt{2\pi \sigma^{2}_{S}}} \exp \left( - \frac{(S^{z} - \mean{S^{z}})^2}{2\sigma^{2}_{S}} \right ) \,\, .
\label{eq:PSz}
\end{eqnarray}
In the above, the variance of $S^{z}$, $\sigma^2_{S}$ is given by:
\begin{equation}
\sigma^{2}_{S} = \mean{(S^{z})^{2}} - \mean{S^{z}}^2 = -k_{B}T \frac{1}{g\mu_{B}} \frac{\partial}{\partial B} \mean{S^{z}} \,\, , \label{eq:sigma}
\end{equation}
which follows from application of the Fluctuation-Dissipation Theorem with $T$ being the Mn spin temperature. The magnetic field dependence of $\mean{S^{z}}$ is given by the well-established, phenomenologically modified Brillouin function,\cite{gaj79}
\begin{equation}
\label{eq:Sz}
\langle S^z \rangle = - N_{\text{Mn}} \cdot S \cdot B_{S}\left(\frac{g \mu B \cdot S}{k_B(T+T_0)}\right)
\end{equation}
where $g$, $\mu_B$, and $k_B$ are the Mn ion $g-$factor, the Bohr magneton, and Boltzmann constant, respectively, and the spin of an individual Mn ion $S=5/2$. The Mn impurity spins in a II-VI material can be considered paramagnetic only in the limit of very small concentration $x$. At $x>0.01$ the short-range antiferromagnetic interactions (both isotropic and Dzyaloshynsky-Moriya anisotropic ones) between the Mn spins have to be taken into account. Since in this work we are interested in QDs with a rather large Mn contents, $N_{\text{Mn}}$ is usually substantially smaller than the number of Mn ions physically present within the QD volume $V$. This is because the Mn spins located on nearest neighbor cation sites are antiferromagnetically coupled with isotropic superexchange (with interaction strength $J \! \approx \! 1$ meV), thus removing these spins from paramagnetic response for fields smaller than $\approx \! 10$ T. 
In the following discussion of optical properties of the \CdMnTe\ QDs, we will only consider these Mn ions, which do not possess a Mn as a nearest neighbor and label them as {\em paramagnetic Mn ions}, which is not strictly correct since next nearest neighbors are also antiferromagnetically coupled and also decrease the paramagnetic response to the magnetic field. Therefore, the number of paramagnetic Mn ions $N_{\text{Mn}}$ will be treated as a model parameter accounting for all the Mn ions responding to fields below 10 T. In the expression (\ref{eq:Sz}), the replacement of temperature $T$ with $T+T_{0}(x)$ accounts for a suppression of paramagnetic response due to the farther-neighbor Mn-Mn couplings. In this way we incorporate the realistic description of interacting Mn (i.e., only approximately paramagnetic) spins into the calculation of various statistical properties of the exciton-Mn complex.\cite{die83prb,bac02,bra02}

\subsection{Description of exchange-coupled system of Mn spins and an exciton} \label{sec:XMn}
The Hamiltonian of the electron-hole pair (the exciton) and the Mn spins is given by:
\begin{equation}
\label{ham}
\hat{H} = \hat{H}_{s-d} + \hat{H}_{p-d} + \hat{H}_{Z} + \hat{H}_{X} \,\, ,
\end{equation}
consisting of the {\em s-d} (electron-Mn) exchange interaction, and the {\em p-d} (hole-Mn) exchange interaction, the Zeeman terms, and the $\hat{H}_{X}$ term related to the electron-hole interaction within the exciton.
The {\em sp-d} exchange interaction is given by \cite{gaj10}
\begin{multline}
\hat{H}_{s-d} + \hat{H}_{p-d} = \\ -\alpha\sum_{i} \delta(\mathbf{r}_{e}-\mathbf{R}_{i})\mathbf{S}_{i}\cdot\mathbf{s} -\beta\sum_{i} \delta(\mathbf{r}_{h}-\mathbf{R}_{i})\mathbf{S}_{i}\cdot\mathbf{j}/3 \,\, , \label{eq:Hspd}
\end{multline}
where $\alpha$ ($\beta$) is the exchange constant for electrons (holes) with $N_{0}\alpha \! = \! 0.22$ eV ($N_{0}\beta \! = \! -0.88$ eV) in \CdMnTe, $\mathbf{r}_{e}$ ($\mathbf{r}_{h}$) is the position operator of an electron (a hole), $\mathbf{s}$ ($\mathbf{j}$) is the spin operator of an electron (a hole), $\mathbf{R}_{i}$ is the position and $\mathbf{S}_{i}$ is the spin operator of the $i$-th Mn ion. $N_{0}$ is the cation density. We consider an exciton in its orbital ground state, i.e., both the electron and the hole occupying respective lowest-energy orbitals, and we assume that its wavefunction is a product of electron and hole wavefunctions. We note that this is only true for strongly confined excitons, where Coulomb correlation effects can be neglected. We obtain the Hamiltonian of the exciton--Mn exchange by projecting $\hat{H}_{X-Mn} = \hat{H}_{s-d} + \hat{H}_{p-d}$ onto two, two-dimensional subspaces corresponding to the electron and the hole. For the electron, we obtain the {\em s-d} exchange as:
\begin{equation}
\label{Hsd}
\hat{H}_{s-d} = -\alpha\sum_{i} |\Psi_{e}(\mathbf{R}_{i})|^{2} \mathbf{S}_{i}\cdot\mathbf{s} \,\, ,
\end{equation}
where  $\Psi_{e}(\mathbf{r})$ is the electron envelope wavefunction. Such a projection leads to a more complicated result in the case of holes. In low-dimensional structures such as QDs, the quantum confinement and strain splits the heavy hole (hh) and light hole (lh) states. In CdTe QDs, the topmost valence band orbital is a twofold spin-degenerate state of mostly hh character, and the projection of the p-d exchange operator on this state gives:\cite{bes04}
\begin{equation}
\label{Hpd}
\hat{H}_{p-d} = -\beta\sum_{i} |\Psi_{h}(\mathbf{R}_{i})|^{2} S^{z}_{i}j^{z}/3 \,\, .
\end{equation}
where  $\Psi_{h}(\mathbf{r})$ is the hole envelope wavefunction.
We note that keeping a finite admixture of lh state in the topmost valence orbital changes the $S^{z}_{i}j^{z}/3$ term to $(S^{z}\kappa^{z} + \epsilon S^{+}\kappa^{-} + \epsilon^{*} S^{-}\kappa^{+})/2$, where $\kappa_{i}$ are the Pauli operators defined in the two-dimensional subspace of hole orbital ground state, and $|\epsilon| \! \ll \! 1$ is a measure of the amount of lh admixture.\cite{leg07,gor10,cyw10} The mixing between heavy and light hole states in CdTe dots is much stronger than in InGaAs dots, resulting in such effects as linear polarization of the trion PL ,\cite{leg07} non-zero in-plane hole $g$-factor and in-plane emission from the dark excitons,\cite{smo12}. However, for the effects considered here it is of minor importance, and in the calculations presented in this paper we will neglect it (i.e.,~we put $\epsilon \! = \! 0$).

For magnetic field along the $z-$axis, the Zeeman term is given by:
\begin{equation}
\hat{H}_{Z} = g_{e}\mu_{B}Bs^{z} + g_{h}\mu_{B}Bj^{z} + g\mu_{B}S^{z} \,\, ,
\end{equation}
where $s^{z}$ is the $z$ component of the electron spin operator, $j^{z}$ is the $z$ component of the hole angular momentum operator, and $S^{z}$ corresponds to the total Mn spin, i.e., $S^{z} \! = \! \sum_{i}S^{z}_{i}$.

The exciton Hamiltonian $\hat{H}_{X}$ includes both isotropic and anisotropic exchange interactions.\cite{bay02} The energy scale of the latter term is $\sim \! 0.1$ meV and thus we neglect it as it is some two orders of magnitude smaller than typical {\em sp-d} exchange energies considered here. The former term is:
\begin{equation}
\label{Hz}
\hat{H}_{X} = \frac{\delta_{0}}{2} \left( \ket{1}\bra{1} + \ket{-1}\bra{-1} - \ket{2}\bra{2} - \ket{-2}\bra{-2} \right )  \,\, ,
\end{equation}
in which $\delta_{0} \! \approx \! 1$ meV is the isotropic e-h exchange splitting, $\ket{\pm 1}$ are the bright exciton states (with electron and hole spins antiparallel) optically active in $\sigma_{\pm}$ polarization, and $\ket{\pm 2}$ are the dark exciton states (with electron and hole spins parallel). The recombination of dark excitons, as mentioned above, is allowed by the mixing between the hh and lh subbands, but this mechanism is of negligible importance in Mn-doped QDs since here the main mechanism of bright-dark exciton mixing is due to flip-flop terms of {\em sp-d} exchange\cite{gor10,cyw10}. Below, we show that the influence of the dark states on the PL line shape of the dots considered here (i.e., containing at least 20 Mn ions) is negligible. Thus, we will neglect $\hat{H}_{X}$ in the following calculations. The only indirect effect of $\hat{H}_{X}$  which we keep, is that together with $\hat{H}_{p-d}$ it suppresses the electron-Mn spin flip-flops (i.e.~the $S^{\pm}_{i}s^{\mp}$ terms in Eq.~(\ref{Hsd})). It allows us to write the exciton--Mn exchange interaction neglecting the off-diagonal terms in $S^{z}$. Thus, we get:
\begin{equation}
\hat{H}_{X-Mn} = -\sum_{i}S^{z}_{i} \left [  \alpha|\Psi_{e}(\mathbf{R}_{i})|^{2}s^{z} + \beta|\Psi_{h}(\mathbf{R}_{i})|^{2}j^{z}/3 \right ] \,\, , \label{eq:HXMn}
\end{equation}

The final approximation that we will make is the assumption that the wavefunctions of both an electron and a hole are of a box (muffin-tin) shape, i.e., the electron and hole wave functions are approximated by a constant value inside a volume $V_{e/h}$: $\Psi_{e,h}(\mathbf{r}) \! = \! 1/\sqrt{V_{e,h}}$, while vanishing outside.
This approximation, although crude, allows for easy diagonalization of $\hat{H}_{X-Mn}$ and was shown to correctly reproduce the static properties of the exciton--Mn system.\cite{woj07,woj08} In the muffin-tin approximation, the exciton--Mn interaction reads:
\begin{equation}
\hat{H}_{X-Mn} = -S^{z} \left (  \frac{\alpha}{V_{e}} s^{z} + \frac{\beta}{V_{h}}j^{z}/3 \right )
\label{eq:HXSz}
\end{equation}

\subsection{Modeling of the PL spectra}
\label{sec:PLmod}
The two distinct confinement volumes for the electron and the hole in Hamiltonian (\ref{eq:HXSz}) enable us to treat separately the confinement of both carriers, {\em a priori} unknown. On the one hand, negligible band offset in the CdTe/ZnTe interface suggests a weakly bound hole, leading to a much stronger leakage of the hole wavefunction out of the dot. On the other hand, PL experiments on CdTe dots with single Mn ions allow to estimate the relative electron and hole confinement and point out that the hole is in fact more strongly bound.\cite{bes07,gor10} The electron confinement volume is usually identified with the volume of the QD. In the following, we define an effective confinement volume $V$. Taking into account the ratio of $|\beta/\alpha| \! = \! 4$ and the realistic bound of $1/2 \leq V_{h}/V_{e} \! \leq \! 2$, one can show that for both cases: (i) Mn spins located only within the QD, and (ii) the same Mn concentration in the barrier and the dot, the effective confinement volume $V$ is to a very good approximation equal to the volume occupied by the hole (see Appendix \ref{app:V} for details). Using the Hamiltonian (\ref{eq:HXSz}) for a given value of $S^{z}$ we obtain the energies of the bright $\ket{\pm 1}$ excitons.
\begin{equation}
E_{\pm}(S^{z}) = \mp S^{z} \frac{1}{2}\left ( \frac{\beta}{V_{h}} - \frac{\alpha}{V_{e}} \right ) \pm \frac{E_{Z}}{2} \equiv  \pm S^{z} \frac{1}{2} \frac{\alpha-\beta}{V} \pm \frac{E_{Z}}{2} \,\, ,  \label{eq:EV}
\end{equation}
where $E_{Z}$ is the Zeeman splitting of the bright exciton states due to applied magnetic field.

The PL intensity in $\sigma_{\pm}$ polarization corresponding to recombination of $\ket{\pm 1}$ excitons at energy $E$ relative to the energy of a bare exciton (noninteracting with the Mn ions, its energy determined solely by the QD morphology) is then given by:
\begin{equation}
I_{\pm}(E) \propto  \int P(S^{z}) p(E_{\pm}(S^{z})) \delta(E-E_{\pm}(S^{z})) \text{d}S^{z}  \,\, , \label{eq:PLSz}
\end{equation}
where $p(E_{\pm})$ is the probability of occupation of exciton $\ket{\pm 1}$ states at the moment of recombination. Note that we assume that the nonresonant excitation leads to creation of both $\ket{\pm 1}$ with equal probability.

In the completely unrelaxed case we have $p(E_{\pm}) \! = \! 1/2$ and the exciton just probes the statistical distribution of the Mn ions. The PL spectrum is then given by:
\begin{equation}
I_{\pm}(E) \propto  \exp \left[ -\frac{(E \mp E_{0})^2}{2\sigma^{2}_{E}} \right ]
\label{eq:PLun}
\end{equation}
with
\begin{equation}
\sigma^{2}_{E} = \frac{1}{4V^{2}}(\alpha-\beta)^2\sigma_{S}^{2} \,\, ,  \label{eq:sigma_E}
\end{equation}
where $\sigma_{S}$ is given by Eq.~(\ref{eq:sigma}), and
\begin{equation}
E_{0} = \frac{\alpha-\beta}{2V}\langle S^{z} \rangle \pm E_{Z} \,\, ,
\label{gze}
\end{equation}
where $\langle S^{z} \rangle$ is given by Eq.~(\ref{eq:Sz}). Note that in a nonzero $B$ field the exciton Zeeman splitting $E_{Z}$ is typically negligible compared to the first term in the above equation, and in the following we neglect the presence of $E_{Z}$ term.

\begin{figure}[!h]
  \includegraphics[angle=0,width=.5\textwidth]{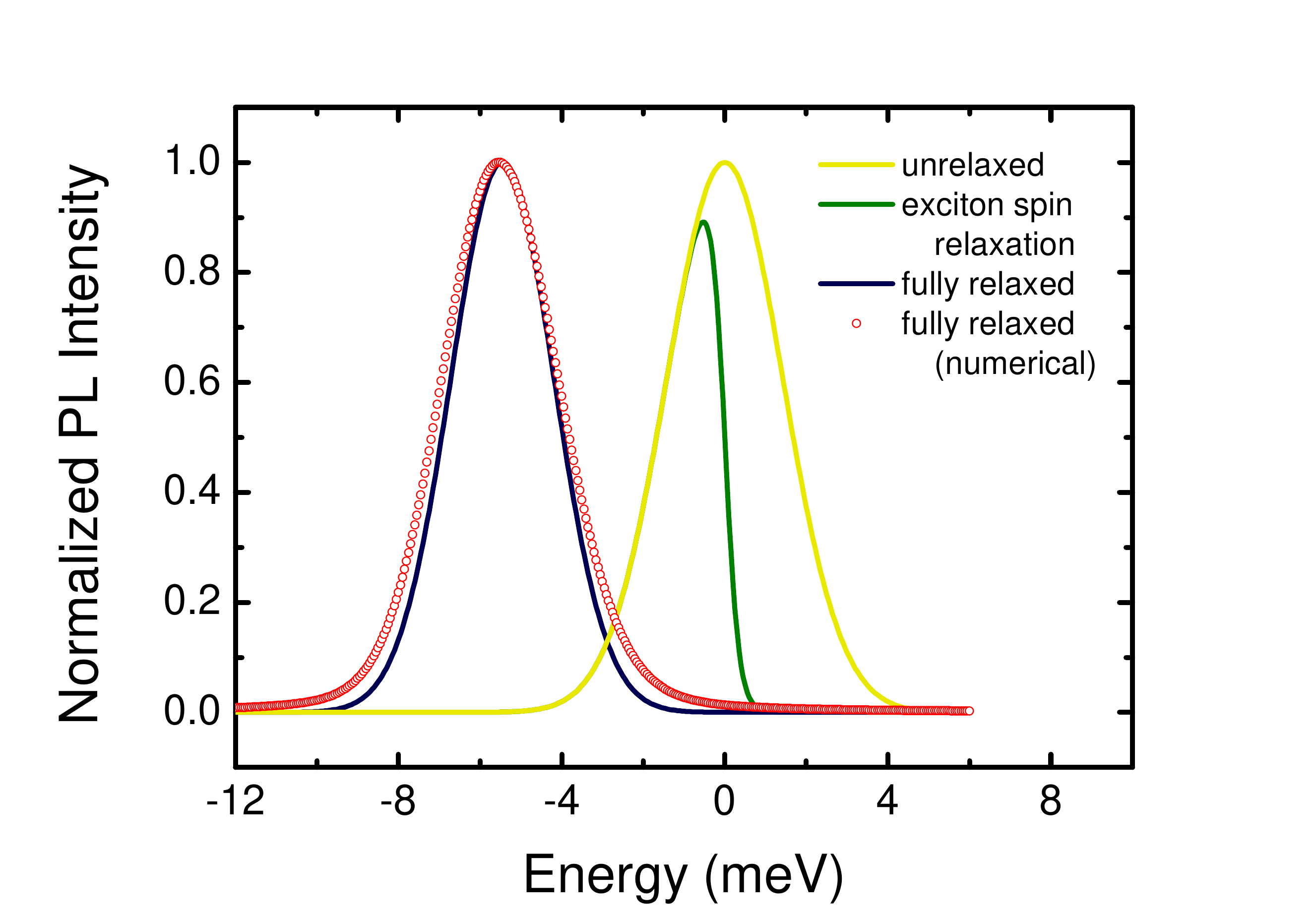}
  \caption{Calculated PL spectra for a QD with an exciton localization volume covering $N_0V = 4000$ cation sites and with $N_{\text{Mn}} = 40$ paramagnetic Mn ions at a temperature of 4 K. Three stages of the relaxation process are modeled. Points demonstrate result of numerical diagonalization of a Hamiltonian, which includes the off diagonal terms, neglected in developing the analytical model presented in the text. Lines denote the results of the analytical model. }
  \label{fig:models}
\end{figure}

\begin{figure}[!h]
  \includegraphics[angle=0,width=.5\textwidth]{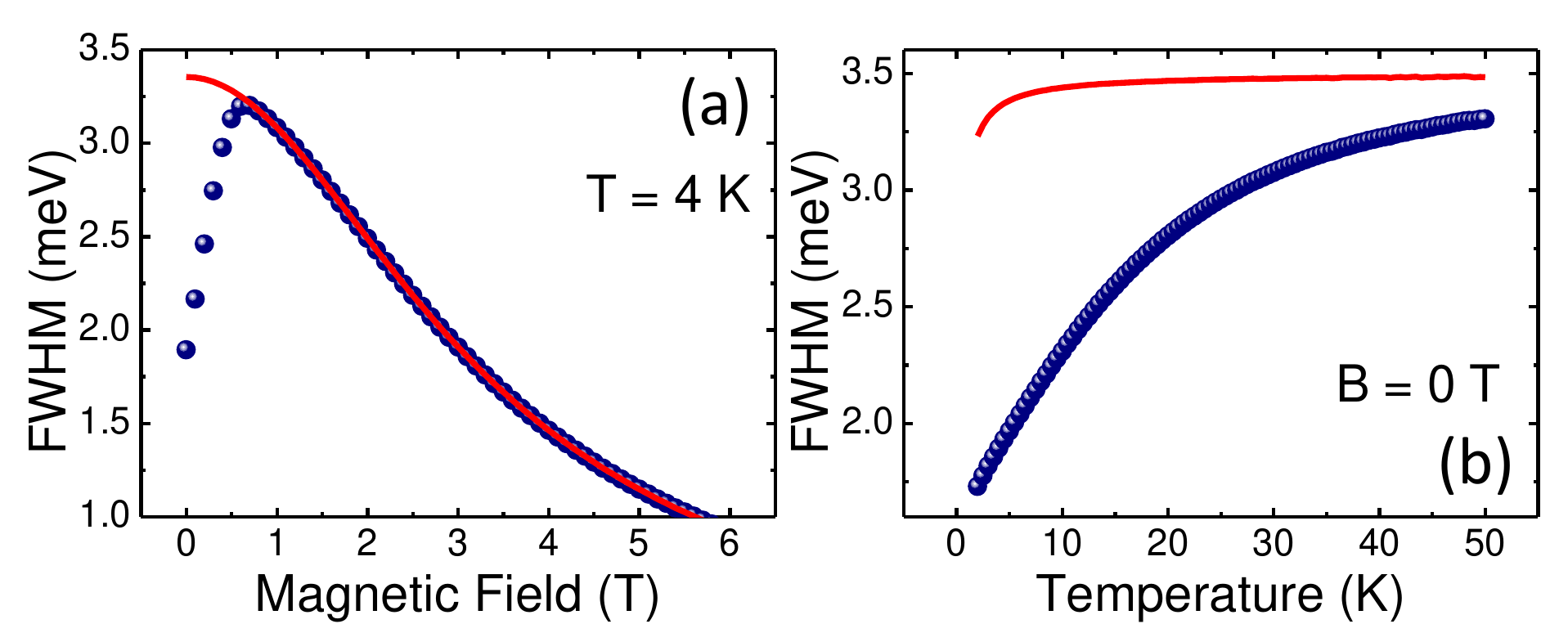}
  \caption{Magnetic field (a) and temperature (b) dependence of calculated transition linewidths (FWHM) for a QD with an exciton localization volume covering $N_0V = 4000$ cation sites and with $N_{\text{Mn}} = 40$ Mn ions. Points (lines) denote the linewidths s of a recombination of fully spin-relaxed (completely unrelaxed) excitons.}
  \label{fig:fwhms}
\end{figure}

We see that in the completely unrelaxed case the line shape is Gaussian with the energy position and FWHM ($\gamma$) determined by the mean and variance of $S_z$, respectively, which in turn are magnetic field dependent. Using the above expressions FWHM of the PL transition is given by:
\begin{equation}
\gamma (B) =  \sqrt{8 \ln 2 \cdot \left( \frac{N_0(\alpha - \beta)}{2N_0V} \right)^2 \frac{k_B T}{g \mu_B} N_{\text{Mn}} S \left( -\frac{\partial B_S}{\partial B}\right)}
\label{gammab}
\end{equation}
where the argument of the Brillouin function $B_S$ is the same as in Eq. \ref{eq:Sz}. In particular, at $B=0$, $\langle S_z \rangle = 0$ and the PL transition is centered around the energy of a bare exciton (see Fig. \ref{fig:models} yellow (light) curve). Naturally, the expression (\ref{eq:PLun}) reproduces both the splitting and the narrowing of the transition line with increasing magnetic field (see line in Fig. \ref{fig:fwhms}(a)).

If the exciton spin is relaxed, the occupation of the $\ket{\pm 1}$ exciton states split by the {\em sp-d} exchange interaction is thermal and the majority spin population at low enough temperature probes roughly a half of the Mn spin states --- those aligned antiparallel with the majority exciton spin. We take this into account by introducing the occupation factors:
\begin{equation}
p(E_{\pm}(S^{z})) = \frac{\exp(\frac{-E_{\pm}(S^{z})}{k_BT_{X}})}{2\cosh(\frac{-E_{\pm}(S^{z})}{k_BT_{X}})}
\end{equation}
where we have again neglected the $E_{Z}$ term. In the above, we have introduced a temperature $T_{X}$, which corresponds to the exciton spin temperature in the limit of the exciton spin relaxation time much shorter than recombination time. Since the exciton relaxation time decreases with increasing Mn concentration,\cite{klo11pol}, this is true only for sufficiently large $x$ and thus in principle $T_X$ represents an upper limit for the exact exciton spin temperature. Indeed, in a time-{\em unresolved} spectrum, we integrate both the recombination of spin-unrelaxed and spin-relaxed excitons. Quantitatively, $T_X$ characterizes the degree to which the exciton spin population has equilibrated with the reservoir to which the exciton spin degree of freedom is most strongly coupled. Using  Eq.~(\ref{eq:PLSz}) we get:
\begin{equation}
I_{\pm}(E) \propto  \exp \left[ -\frac{(E \mp E_{0})^2}{2\sigma^{2}_{E}} \right ] \frac{\exp(\frac{-E}{k_BT_{X}})}{2\cosh(\frac{E}{k_BT_{X}})}
\label{PLXrel}
\end{equation}
An immediate consequence of this expression is that with increasing positive magnetic field ($B>0$ leading to $\langle S^{z} \rangle <0$ and $E_{0} < 0$) we obtain a redshift of $\sigma_{+}$ emission, while the $\sigma_{-}$ emission is blueshifted, and its intensity drops rapidly to zero once $|E_{0}/k_{B}T_{X}|$ becomes larger than one. However, a more interesting consequence is that in the case of relaxed exciton spin, at zero magnetic field, the PL line shape is no longer Gaussian, but becomes suppressed from the high energy side due to the Bolztmann  factor (see Fig. \ref{fig:models} green (dark) curve). With decreasing the temperature, the asymmetry is enhanced as a result of larger imbalance in the thermal occupation of the exciton spin states (see Fig. \ref{fig:fwhms}(b)). On the other hand, as the magnetic field is increased, the exchange-driven spin splitting of the exciton states increases and the occupation probability of the spin-up state quickly approaches unity. Thus, the PL lines recover the Gaussian shapes for magnetic fields of about 1 T depending on temperature (see Fig. \ref{fig:fwhms}(a)).

In order to model the PL spectrum for the exciton--Mn spin system in full thermal equilibrium, we have to take into account the formation of the polaron (see Introduction). The polarization of the Mn ions occurs as a result of an effective magnetic field $B_{ex}$ imposed by the exchange interaction with the exciton:
\begin{equation}
B_{ex}  = \frac{(\alpha-\beta)}{2 g \mu_B V} \,\, .
\label{Bex}
\end{equation}
Thus, once the Mn population achieves a thermal equilibrium in this field, a net magnetization is created, and a redshift of the PL peak is developed. Importantly, at external magnetic field $B_{0}\! = \! 0$ the orientation of the developed magnetization is random, as it depends on the random initial magnetization of fluctuating Mn spin ensemble,\cite{klo11pol} resulting in no net polarization of PL from the redshifted peak (unless a memory effect is present\cite{gur08}). With increasing $B_{0}$, the PL rapidly becomes $\sigma_{+}$-polarized. The polarization-unresolved PL signal in the regime of polaron formation is thus given by Eq.~(\ref{PLXrel}) albeit with $E_{0}$ and $\sigma_{E}$ calculated  at the value of the {\em total field} of $B \! = \! B_{0} + B_{ex}$.\cite{mak00} Since the effective field $B_{ex}$ is on the order of a few tesla leading to a redshift of the PL peak exceeding $\sigma_{E}$, we expect the lineshape for a fully equilibrated system to be a Gaussian.


In Fig. \ref{fig:models} we demonstrate the results of our model calculations for a dot confining an exciton to a volume spanning over $N_0V = 4000$ cation sites with $N_{\text{Mn}} = 40$ Mn ions at a temperature of 4 K. Points denote the PL spectrum obtained numerically by diagonalization of the Hamiltonian, which includes the off-diagonal spin-flip terms\cite{woj07,woj08} and assumes full thermal equilibrium. This result is compared to the spectrum (dark blue line) calculated with an analytical formula (\ref{eq:PLun}) at $B_{ex} = 1.2$ T, which corresponds to the assumed localization volume $V$ (see Eq. (\ref{Bex})). We note that the agreement, despite dropping the off-diagonal terms in developing expression (\ref{eq:PLun}), is very good. Also, for a fully thermalized exciton--Mn ions system, the spectrum is symmetric and Gaussian, as expected. The redshift of the exciton energy directly reflects the gain of the energy of the system upon the build-up of the magnetization. In this case, the mean exciton energy is more than 5 meV lower than for a dot where the full equilibrium was not reached. The spectrum calculated for the unrelaxed case is just a Gaussian centered at zero --- the energy of a bare, noninteracting exciton. However, for the spin-relaxed case, we find that the spectrum is strongly asymmetric and the linewidth is substantially reduced with respect to both the unrelaxed and the fully relaxed case. This is a strong indication that in the case of spin-relaxed excitons, the linewidth dependence on magnetic field can be non-monotonic. Indeed, as shown in Fig. \ref{fig:fwhms}(a), the PL transition linewidth considered in Fig. \ref{fig:models} increases with the magnetic field up to 1 T and then decreases and follows the dependence for unrelaxed excitons, i.e., directly monitors the magnetization fluctuations, which are then suppressed with increasing the magnetic field. In Fig. \ref{fig:fwhms}(b), we compare the temperature dependence of the linewidth for the unrelaxed and spin-relaxed cases. It demonstrates that the narrowing due to exciton spin relaxation is temperature dependent and that at liquid helium temperatures the linewidth can be twice smaller than in the unrelaxed case.

\section{Samples and Experiment}
\label{samexp}

The samples are grown by molecular beam epitaxy on (100)-oriented GaAs substrates. First, a 4 $\mu$m-thick CdTe buffer layer is deposited. Then, the first barrier layer, about 0.6 $\mu$m thick is grown. \CdMnTe\ dots are formed on top of this layer and covered with another barrier layer, 50 nm thick. QDs are formed using the tellurium desorption method \cite{tin03,woj10} from a 6 monolayer-thick \CdMnTe\ layer. Dots with $x\leq0.035$ are formed in \CdZnTe\ barriers with $y\approx 0.7$. On the other hand, dots with $x>0.035$ are formed in \ZnMnTe\ barriers with Mn concentration $x$ kept nearly uniform in the barrier layers and in the dots. Such procedure ensures that \CdMnTe\ QDs with a high Mn content maintain type-I band alignement as opposed to embedding the dots in ZnTe. Indeed, incorporation of a large density of Mn ions into the \CdMnTe\ dot leads to an increase of the QD band gap. Since the valence band offset in a CdTe/ZnTe heterojunction is negligible, most of the hole confinement results from strain-induced band shifts. Increasing the Mn content decreases the lattice mismatch between \CdMnTe\ and ZnTe and thus diminishes strain. As a result, above a certain value of $x$ the dot potential for holes can become repulsive. Moreover this procedure excludes any effects related to interface interdiffusion \cite{gri96} and ensures a homogenous distribution of Mn ions in the dots. In the following, we refer to particular samples by their nominal Mn content $x$.

We measure cw or time-resolved photoluminescence (PL) of single QDs excited with a 532 nm solid state laser or with a frequency doubled output beam of a optical parametric oscillator pumped with a Ti:Sapphire picosecond pulsed laser, respectively. Laser beam is focused with a large numerical aperture microscope objective. In order to access single dots, apertures with diameters of 500 nm are produced by spin-casting polystyrene beads and evaporating a 200 nm thick gold mask. The beads are subsequently lift-off by rinsing the sample in trichloroethylene and methanol. PL signal is collected with the same objective and analyzed with a CCD camera and a monochromator. Time-resolved detection is performed with a synchroscan streak camera with a temporal resolution of about 15 ps. For zero magnetic field measurements the sample is kept in a cold finger cryostat at 10 K. For the studies of the Zeeman effect, the sample is immersed in pumped helium at 2 K in a split-coil cryostat providing fields up to 6 T in Faraday configuration. Two circular polarizations of the emitted light are recorded.

\section{Experimental Results}
\label{expres}

\subsection{Three relaxation stages}

In a previous report \cite{klo11pol}, we demonstrated that the equilibrium between the Mn ions and the exciton (i.e., the EMP formation) is reached after the exciton has relaxed its spin and that establishing of the equilibrium during the exciton lifetime requires a concentration of Mn ions $x\geq0.1$. Thus, for a sufficiently high $x$, the system is completely unrelaxed right after the photoexcitation, then the exciton adjusts its spin and then the EMP is formed. If the Mn density is not high enough, the exciton relaxes its spin, but the formation of the polaron is interrupted by recombination. In Section \ref{sec:PLmod}, we showed that each of the three relaxation stages leads to a particular shape of the PL spectrum. Now, we compare this prediction with the experiment.

\begin{figure}[!h]
  \includegraphics[angle=0,width=.5\textwidth]{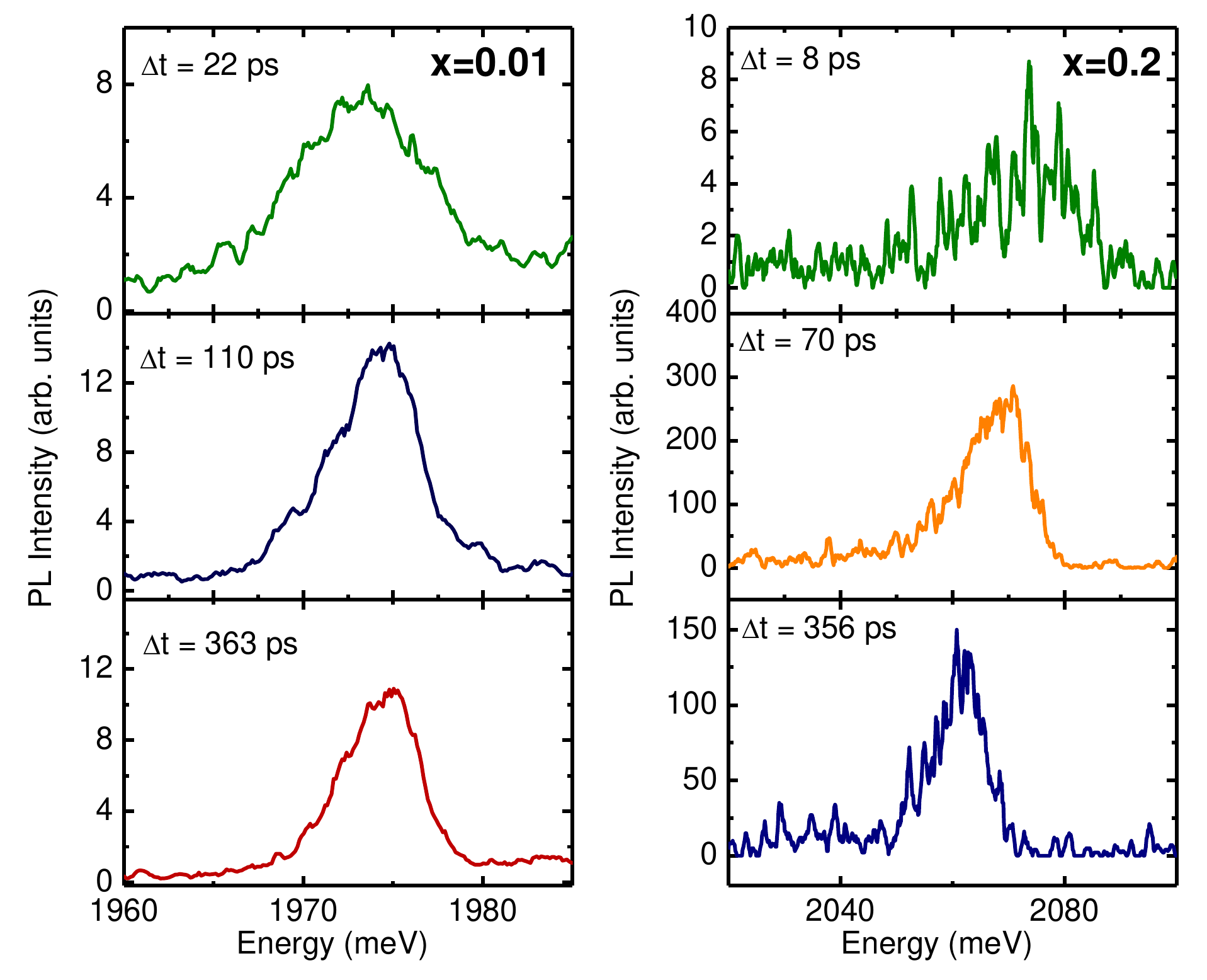}
  \caption{Transient PL spectra for a QD with $x=0.01$ (integrated over 15 ps, left panel) and $x=0.2$ (integrated over 8 ps, right panel) for three time delays $\Delta t$ after the excitation pulse.}
  \label{fig:tres}
\end{figure}

In Fig. \ref{fig:tres}, we show transient PL spectra at different time delays after the excitation pulse. For the sample with $x=0.01$ (left panel), the spectrum is broad shortly after the excitation (top spectrum) and with increasing delay it narrows and becomes asymmetric without changing much the central energy position (middle and bottom spectra). The spectrum is noticeably steeper on the high than on the low energy side. This is a clear indication of the exciton spin relaxation (confront Fig. \ref{fig:models}). The complete equilibrium is not reached since the expected EMP formation time for this Mn density is in the nanosecond range.\cite{die95,klo11pol} For the sample with $x=0.2$ (right panel), exciton spin relaxation is so efficient that the initial spectra recorded only 8 ps after photoexcitation (top panel) already exhibit a certain degree of asymmetry. It becomes even more clear as we look at larger delays (middle panel). Simultaneously, the spectrum undergoes a redshift reflecting lowering of the exciton energy upon EMP formation, which for this QD develops in $\tau_f=100$ ps.  After reaching the equilibrium, the Mn ions are polarized giving rise to a magnetization resulting from the effective field $B_{ex}$ imposed by the exciton. Consequently, the spectrum becomes symmetric and Gaussian with its FWHM reflecting the magnetization fluctuations within the polaron. Here, we do not attempt fitting of the PL line shapes with Eqs. (\ref{eq:PLun}) and (\ref{PLXrel}), since these spectra do not pertain to any steady states, but rather reflect mixed stages of a dynamical process. However, it is clear that the distinction of specific relaxation stages is correct.

\begin{figure}[!h]
  \includegraphics[angle=0,width=.49\textwidth]{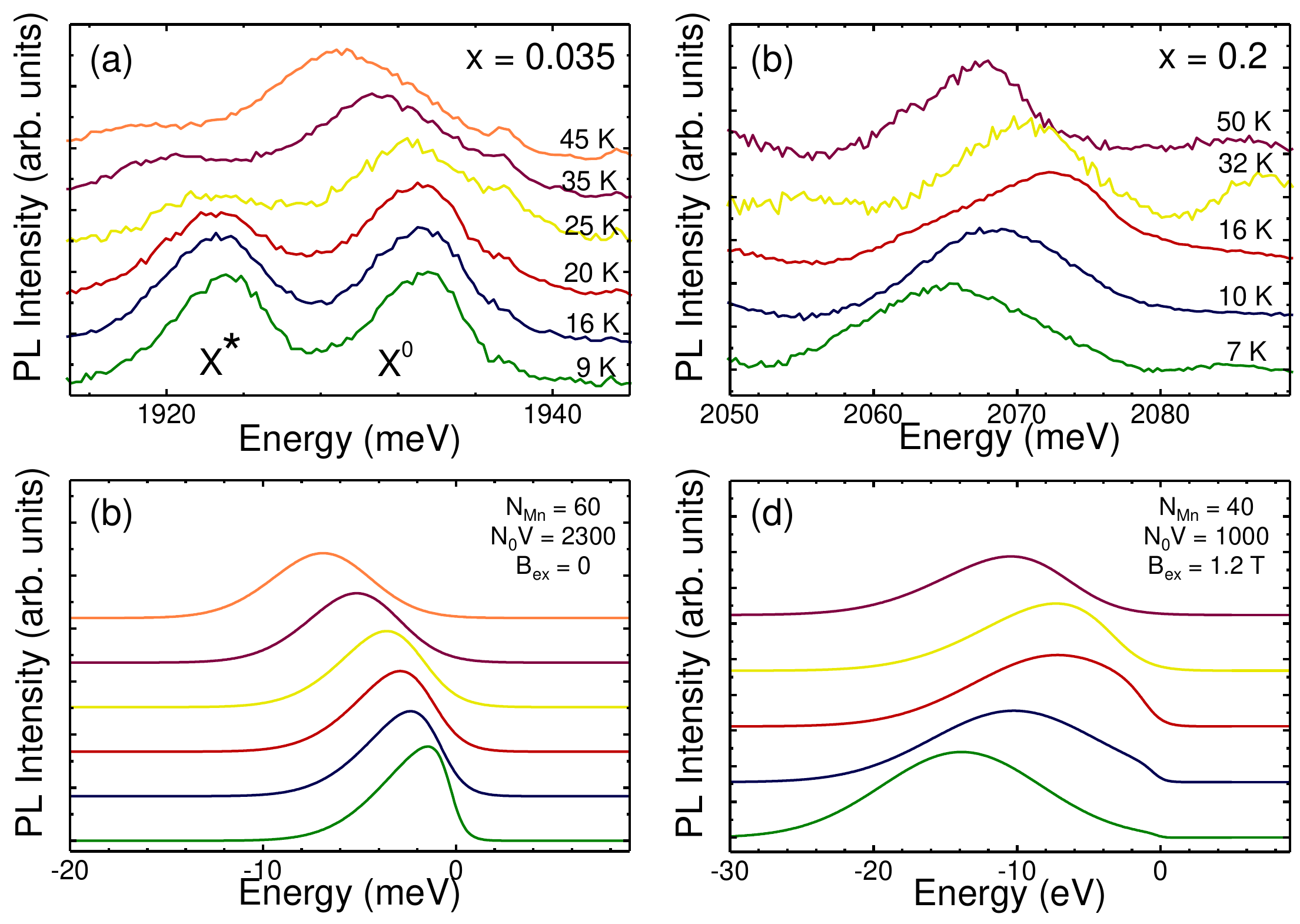}
  \caption{PL spectra measured for a QD with $x=0.035$ (a) and $x=0.2$ (b) at various temperatures compared with calculated ones in, respectively, (c) and (d). For the calculation in (c), spin relaxed excitons are assumed. For the calculation in (d), we assume the EMP formation resulting from the effective field $B_{ex}$. The EMP is suppressed with increasing temperature. The spectra are calculated for the same temperatures as in the experiment and coded with the same colors.}
  \label{fig:tempe}
\end{figure}

Another qualitative agreement between our line shape theory and the experiment is demonstrated in Fig. \ref{fig:tempe}, where we compare measured and calculated PL spectra at various temperatures for two dots: with $x=0.035$ (left panel) and with $x=0.2$ (right panel). In the former case, we observe two transitions, one related to the neutral exciton ($X^0$) and the other to a charged exciton ($X^*$), both of them asymmetric as a consequence of the spin-relaxed exciton population. For this Mn concentration, the exciton spin relaxation time is on the order of 10 ps \cite{klo11pol} --- much shorter than the recombination time, which amounts to about 300 ps.\cite{klo11} We thus observe that these spectra reflect a steady state with a relaxed exciton spin. As the temperature is increased, the imbalance between occupation probabilities of the spin-up and spin-down excitons states becomes less pronounced and the PL spectrum evolves into a symmetric Gaussian (see Eq.~\ref{PLXrel}). Simultaneously, the transitions red-shift as a result of the shrinkage of the band gap. On the other hand, the spectrum of the dot with $x=0.2$ at low temperature is approximately a Gaussian as a result of the total equilibrium attained here during the exciton lifetime (see Fig. \ref{fig:tres} and the discussion above). Note that in these time-integrated spectra an additional broadening due to the transient energy shift reflecting the EMP formation can be present. As the temperature is increased, the magnetization of the EMP is diminished and consequently the exciton recombination energy is blueshifted. However, the excitons remain spin-relaxed and thus with increasing the temperature the PL spectrum {\em acquires} an asymmetry. As the temperature is further increased the spectrum broadens, redshifts, and becomes symmetric as for the $x=0.035$ sample.

\begin{figure}[!h]
  \includegraphics[angle=0,width=.475\textwidth]{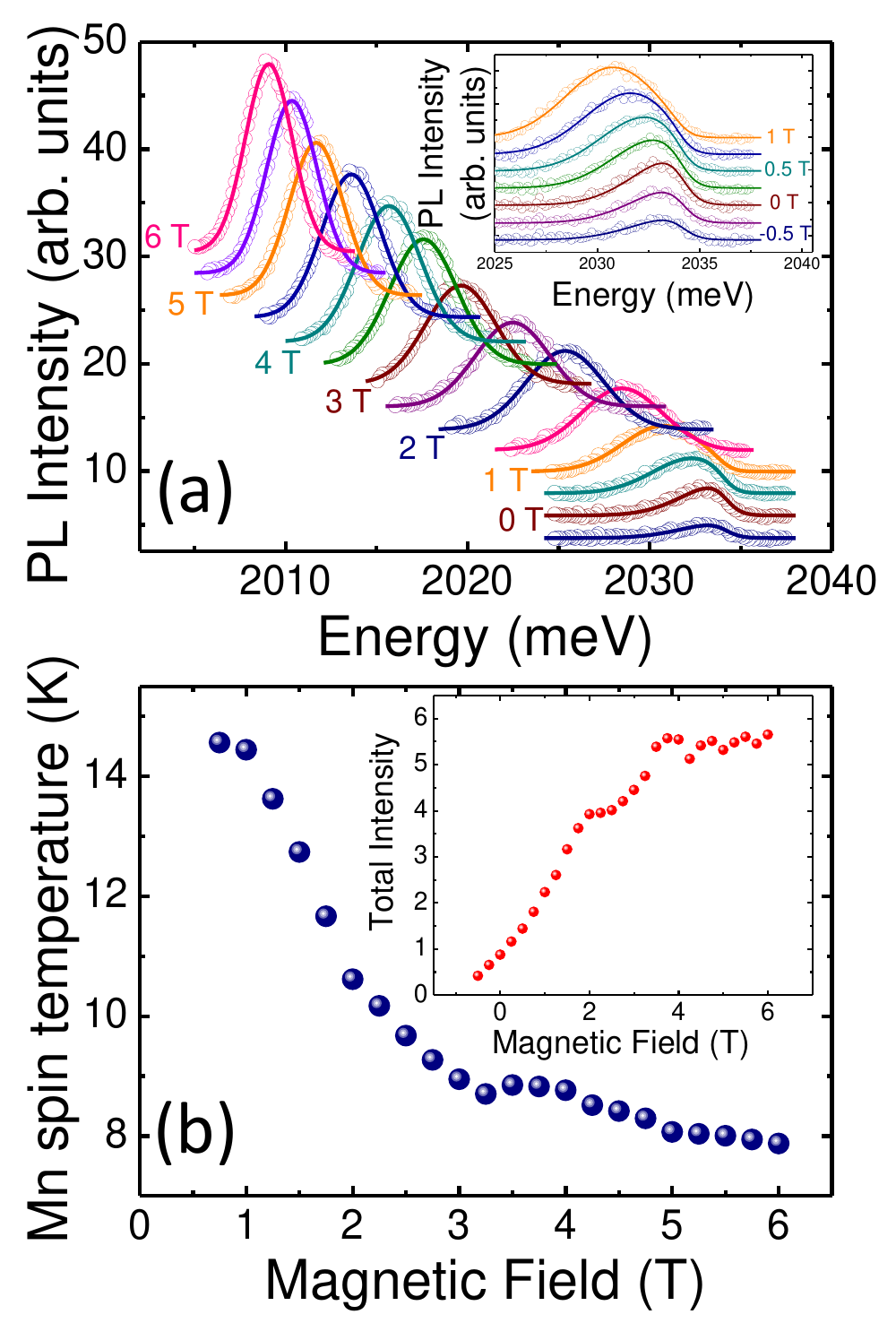}
  \caption{(a) Measured (points) and fitted (lines) PL spectra for a QD with $x=0.035$. The fitting is performed for the entire set of spectra simultaneously yielding parameters $N_{\text{Mn}} =$ 99, $N_0V = 3200$. The inset shows a close-up of the small field spectra. (b) Magnetic field dependent Mn spin temperature obtained from the fitting. Inset:  the fitted, magnetic field dependent integrated intensity.}
  \label{fig:mf-spectra}
\end{figure}

A direct comparison between our transition line model and the experiment is shown in Fig. \ref{fig:mf-spectra}(a), where we present PL spectra for a QD with $x=0.035$ immersed in liquid helium at 2 K, in magnetic fields up to 6 T, positive (negative) sign denoting $\spl$ ($\smn$) polarization of detection. The inset shows the close-up of the spectra for the smallest fields, clearly demonstrating the line shape anisotropy. The asymmetry of the spectrum is more pronounced than for the spectrum for a dot with the same Mn content shown in Fig. \ref{fig:tempe}, where lowest bath temperature was 9 K.
Above 0.5 T, the signal in $\smn$ polarization disappears reflecting the vanishing of the occupation factors introduced in Eq.~(\ref{PLXrel}). 
Points and lines denote the measured and fitted spectra, respectively. In the fitting, we assume completely spin relaxed excitons (see expression (\ref{PLXrel})). The fitting is performed {\em simultaneously} to the whole set of 27 spectra from -0.5 to +6 T with a step of 0.25 T (for the sake of visibility, in Fig. \ref{fig:mf-spectra} we show every 2nd spectrum). We keep constant the model parameters unaffected by the magnetic field, i.e., $N_{\text{Mn}}$, $N_0V$, and $T_X$, while letting the transition amplitude and Mn spin temperature evolve with the magnetic field.

The accuracy of the fit is rather excellent. Assuming a lens shape QD, the obtained localization volume $N_0V = 3985$ corresponds to a dot with 19 nm in diameter and 2 nm in height, which is a typical size.\cite{woj08} The fitted number of paramagnetic Mn ions $N_{\text{Mn}} = 99$ should be analyzed carefully, since this model parameter not only includes those ions which do not have a Mn as a nearest neighbor, but also a combined effect of further neighbors leading to a suppressed paramagnetic response (see discussion in Section \ref{sec:Mn_distribution}). Moreover, statistical fluctuations of the Mn number have to be considered. Following the discussion presented in Section \ref{sec:Mn_distribution}, we can calculate the number of Mn ions without a Mn as a nearest neighbor. For dots with $x=0.035$ grown on \CdZnTe\, the effective Mn concentration $x_{\text{eff}} = 0.024$ and thus, in the studied dot we expect to have $N_0V\cdot x_{\text{eff}} = 96$ paramagnetic Mn ions. We thus find the fitted value in excellent agreement with this expectation, since the number fluctuations are on the order of $\sqrt{N_{\text{Mn}}} \approx $ 10.

In Fig. \ref{fig:mf-spectra}(b), we show the fitted, magnetic field dependent Mn spin temperatures. For magnetic fields smaller than 0.5 T, the Mn spin system is efficiently heated by the energy transfer from the hot, photocreated carriers by simultaneous carrier--ion spin flip-flops.\cite{kon00,cle10} In that case $T_X$ controls the PL lineshape and the Mn spin temperature becomes irrelevant as long as its value is higher than about 10 K. As the magnetic field is increased, a cooling process occurs as the flip-flops become energetically forbidden when the exchange driven carrier Zeeman splitting becomes larger than Mn spin splitting governed by the Mn $g$-factor of 2.\cite{kon00,cle10} As a consequence, the Mn spin temperatures decrease from about 15 down to 8 K. For magnetic fields above about 3 T, the spin temperature is stabilized as the Mn ions attain a thermal equilibrium with the lattice. We thus conclude that the lattice temperature is about 8 K.  It remains a few kelvin larger than the bath temperature of 2 K. We expect that this elevated temperature originates from diffusive heat transport via the LO phonons cascaded down in the energy relaxation process of nonresonantly excited carriers.\cite{hun05} The fitted value of $T_X = 9.9 K$ is slightly higher than the lattice temperature, which is a consequence of the time-integrated nature of this PL measurement (see discussion in Sec.~\ref{sec:PLmod}). In the inset to Fig. \ref{fig:mf-spectra}(b), we plot the total transition intensities as a function of the magnetic field obtained by integrating the fitted spectra. We observe an increase by about a factor of 10 between 0 and 4 T. We attribute this effect to an increased PL yield resulting from blocking of the excitation transfer from the QDs to the internal Mn transitions.\cite{kim00,hun04,naw95}

\subsection{Transition energy and linewidth}

\begin{figure*}
  \includegraphics[angle=0,width=\textwidth]{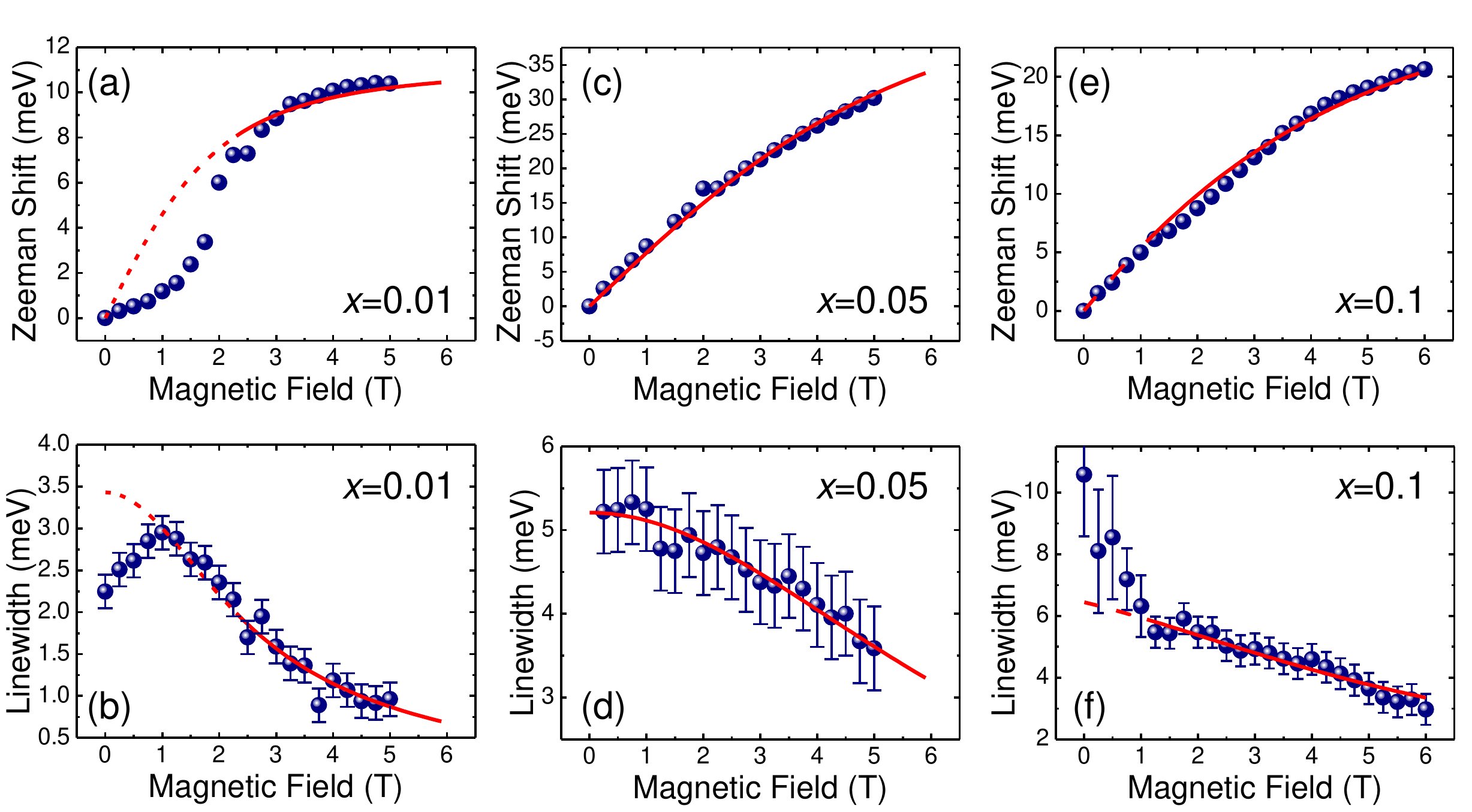}
  \caption{Zeeman shifts (top) and linewidths (bottom) for exciton transitions from \CdMnTe\ QDs with intended Mn contents of $x=0.01$ (a,b) $x=0.05$ (c,d), and $x=0.1$ (e,f). Points are experimental values, lines are Zeeman shifts and linewidths fitted according to Eqs. (\ref{gze}) and (\ref{gammab}), respectively. Fitting is performed {\em simultaneously} for the energy shift and linewidth for each particular dot. For the dots with $x=0.01$ and with $x=0.1$, the data points for low magnetic fields are skipped from the fitting procedure (see dashed lines and the discussion in the text)}.  \label{fig:mf}
\end{figure*}

In this Section, we analyze the transition energy (energy of the PL maximum) and the transition linewidth (FWHM) as a function of the magnetic field for \CdMnTe\ QDs with $x$ ranging from 0.01 to 0.2. The objective of this analysis is to demonstrate how the spectroscopic features change with the Mn content, to retrieve the morphological parameters from the PL spectra, and to discuss what parameters control the particular features at different Mn densities.

In Fig. \ref{fig:mf}, we show excitonic Zeeman shifts (top) and linewidths (bottom) for three \CdMnTe\ dots: with nominal $x=0.01$ (Fig. \ref{fig:mf}(a,b)), with $x=0.05$ (Fig. \ref{fig:mf}(c,d)), and with $x=0.1$ (Fig. \ref{fig:mf}(e,f)). Measured Zeeman shifts are fitted with Brillouin functions according to Eq. (\ref{gze}). The linewidths are fitted with Eq. (\ref{gammab}). For a given dot, the fitting is performed {\em simultaneously} for the two sets of data. Analogous fits are obtained for the two remaining samples, with $x=0.035$ and $x=0.02$. The fitting parameters are $N_{\text{Mn}}$, $N_0V$, and Mn spin temperature $T$. This temperature is assumed here as field independent
and thus the fitting range for samples with $x<0.05$ are adjusted by omitting the data points below 2 T, where heating of the Mn spins occurs, as discussed above. Moreover, an anomalous non-monotonicity of the transition linewidth
is observed for dots with smallest Mn concentrations
(see Fig. \ref{fig:mf}(b)) which {\em cannot} be accounted for by the heating effects. For samples with $x\geq0.05$ we do not expect any significant variations of the Mn spin temperature with the magnetic field, since for such large Mn densities the heat transferred from the hot photocarriers is efficiently dissipated via the spin-lattice relaxation, which is enhanced in the presence of stronger spin-spin interactions.\cite{die95,yak96} However, for the samples in which the complete EMP formation occurs during the exciton lifetime ($x=0.1$ and $x=0.2$), the transition redshift significantly affects the time-integrated PL linewidth (see data points below 1 T in Fig. \ref{fig:mf}(f)), and hence data points below 1 T are omitted. For these samples, we fit the functions from Eqs. (\ref{gze}) and (\ref{gammab}) assuming that the Mn ions experience the total field $B = B_0 + B_{ex}$, where $B_0$ is the external magnetic field and $B_{ex}$ is given by Eq. (\ref{Bex}).

Fit accuracy is very good. Obtained fitting parameters for the samples identified by $x$ are collected in Table I. It shows that the number of paramagnetic ions in a dot can be as high as about 200 provided that the dot volume is large enough. Above $x=0.05$, the $N_{\text{Mn}}$ decreases. This is expected since the concentration of Mn cations without a Mn as a nearest neighbor
also decreases
. Obtained values of the exciton localization volume are expressed in the number of cation sites and vary from about 1000 to 5500. These numbers agree very well with the QD sizes obtained in the atomic force microscopy (AFM) study of an uncapped layer of \CdMnTe\ dots with $x=0.03$ grown on \CdZnTe\ with $y\approx 0.8$. For lens shaped QDs with a height of 2 nm, our range of volumes correspond to diameters from about 10 to 22 nm. For the dots with the largest Mn density of $x = 0.2$, the fitted localization volume is particularly small, which could be due to exciton autolocalization by the spatially inhomogenous EMP potential.\cite{gaj10,yak96} Obtained Mn spin temperatures are again a few kelvin higher than the bath temperature and as above, we conclude that this originates from heating of the lattice itself by the nonresonant excitation. In Table I, we also compare the obtained values of  $N_{\text{Mn}}$ with those calculated $N^{\text{C}}_{\text{Mn}}=N_0V x_{\text{eff}}$, where $N_0V$ is the fitted localization volumes and $x_{\text{eff}}= x S_0(x)/S$, where $S_0(x)$ is the effective spin saturation value applied to parameterize the Brillouin function \cite{gaj10,gri96}. The agreement is very good for smallest Mn concentrations and in all cases the calculated values lie within three standard deviations from the fitted ones.


\begin{table}[!h]
\begin{ruledtabular}
\begin{tabular}{ccccc}
  \noalign{\smallskip} 
  $x$ & $N_{\text{Mn}}$ & $N_0V$ & $T$ & $N^{\text{C}}_{\text{Mn}}$  \\
\hline \noalign{\smallskip}
  0.01   & 24  & 2970 & 3.1 K &  26 \\
  0.035  & 41  & 2040 & 4.2 K & 46   \\
  0.05   & 183 & 5270 & 8.0 K & 150   \\
  0.1    & 69  & 2230 & 7.0 K & 89   \\
  0.2    & 62  & 1120 & 5.4 K & 48  \\
\end{tabular}
\end{ruledtabular}
\label{params}
\caption{The parameters obtained by simultaneous fitting of the transition energy and linewidth for five QD with different nominal Mn compositions. Exciton localization volume $N_0V$ is expressed in cation sites. Fitted numbers of paramagnetic Mn ions are compared with the ones calculated ($N^{\text{C}}_{\text{Mn}}$) assuming a suppression of paramagnetic response by the nearest neighbors.}
\end{table}

\begin{figure}[!h]
  \includegraphics[angle=0,width=.5\textwidth]{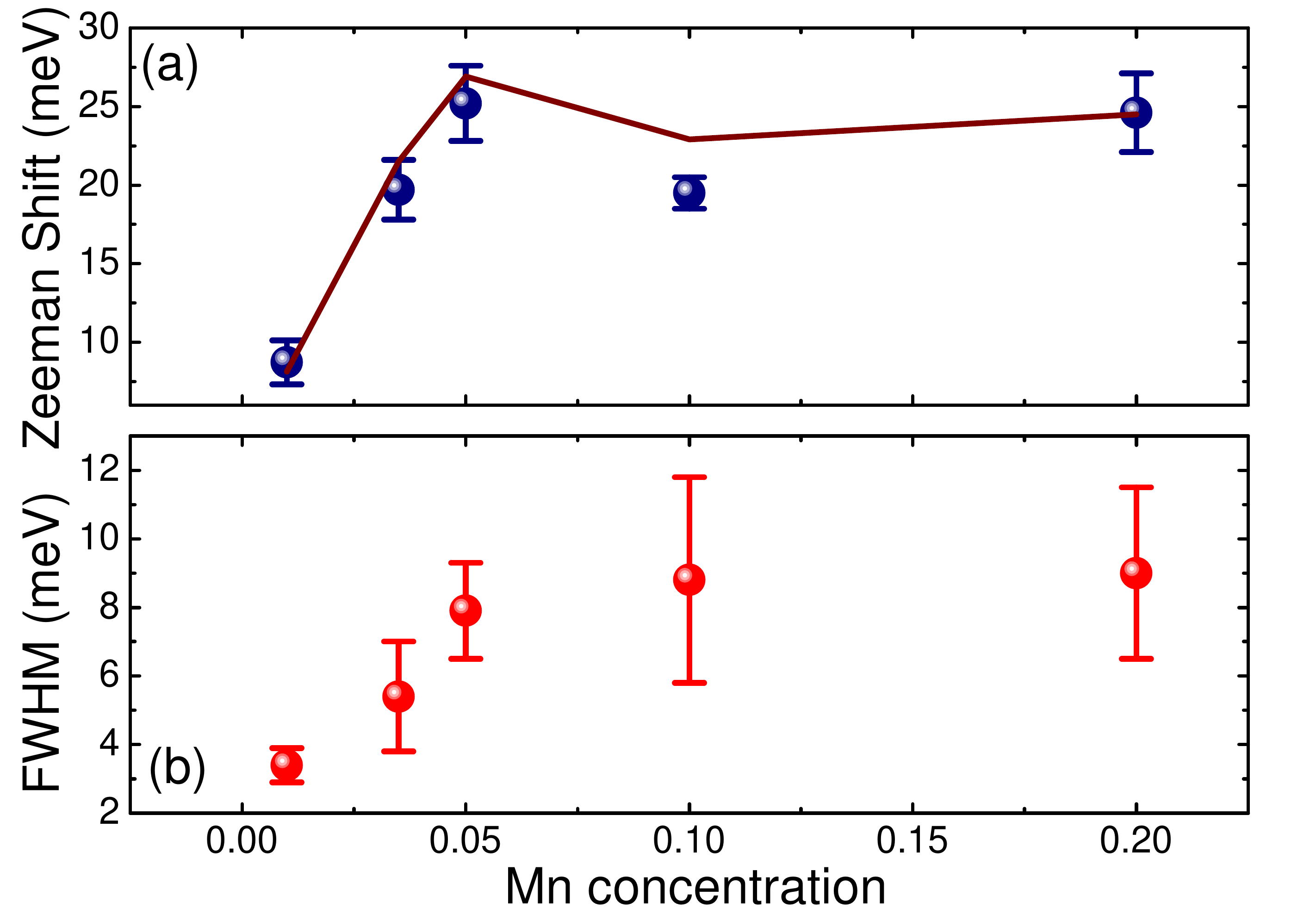}
  \caption{(a) Exciton Zeeman shifts and (b) transition linewidths averaged over a set of dots with different Mn content $x$. Broken line in (a) shows calculated shifts assuming a statistical distribution of Mn ions (see text).}
  \label{fig:FWHM-trene}
\end{figure}

In Fig. \ref{fig:FWHM-trene}(a), we demonstrate Zeeman shifts (points) measured at 5 T with the sample at bath temperature of 2 K averaged over 3-10 dots for each Mn concentration $x$. These values are compared (brown broken line) to Zeeman shifts calculated using Eq. (\ref{gze}) for $T=7$ K and replacing the ratio $N_{\text{Mn}}/N_0V$ by $x_{\text{eff}}$ calculated 
above. Analogously to fitting performed in Fig. \ref{fig:mf}, for $x>0.05$ in the calculation we put $B=5\text{ T}+B_{ex}$ accounting for the EMP formation and $B_{ex}$ is calculated for an average exciton confinement volume spanning over 2000 cation sites. Taking into account that the experimental values correspond to a rather small statistical ensemble, the average values put into the Zeeman shifts calculation reproduce the overall trend very well.

Fig. \ref{fig:FWHM-trene}(b) shows transition linewidths averaged over sets of 10-20 dots for each of the samples with different Mn contents, measured at bath temperature of 10 K and zero external magnetic field. The linewidth monotonically increases with increasing Mn concentration. However, this increase is driven by different mechanisms, depending on $x$. For dots with $x\leq 0.035$, narrow linewidths result from the narrowing due to exciton spin relaxation despite increased Mn spin temperature. On the other hand, for dots with $x>0.035$ the transitions are broadened due to the transient shifts reflecting the EMP formation (still influencing the $x<0.1$ dots even if the full equilibrium is not attained during the exciton lifetime). Hence, the increase of transition linewidth cannot be directly linked to the enhanced magnetization fluctuations, since at different Mn concentrations various mechanisms influence its value. In fact, Eq. \ref{gammab} shows that the linewidth resulting from the magnetization fluctuations scales as $\sqrt{N_{\text{Mn}}}/(N_0V)^2$ and thus for dots at the same stage of the equilibration in the exciton--Mn spin system, we would expect the linewidth to strongly decrease for dots with highest Mn contents.

\section{Discussion}

Experimental results presented in Figs. \ref{fig:tres}, \ref{fig:tempe}, and \ref{fig:mf-spectra} unambiguously demonstrate spin relaxation of {\em excitons} in the studied \CdMnTe\ QDs. This is an important conclusion, since a vast majority of the studies of spin relaxation in QDs focuses on spin relaxation of {\em single carriers}. A natural question arises about the mechanism of the exciton spin relaxation. We propose two types of mechanisms, differing in the role played by the {\em sp-d} exchange interaction.

In the first mechanism, the role of the Mn ions is static and the role of the thermal reservoir is played by phonons. The exchange interaction provides a splitting of the $\ket{\pm 1}$ exciton states. Momentary fluctuations of the QD magnetization results in exchange splittings, which in turn influence the transition linewidth of a time-integrated spectrum. The exchange splitting can be approximated from the transition linewidth taking into account the line narrowing due to exciton spin relaxation or broadening due to transient energy shift during EMP formation. E.g., from the PL spectra for a dot with $x=0.035$ presented in Fig. \ref{fig:mf-spectra}, we can estimate the exchange induced splitting of the exciton state to be equivalent to a direct Zeeman splitting of an exciton in CdTe dot in a magnetic field of about 15 T. In order for the spin-flip transition due to phonon scattering to occur, the Zeeman split sublevels need to have a mixed spin character.\cite{kha01,han07} For non-magnetic QDs, the intermixing can be provided by e.g., the combined effect of the strain-driven heavy-light hole mixing and the electron-hole exchange interaction. The mixing allows a phonon-assisted spin-flip transition and the resulting relaxation rates between the Zeeman split bright exciton states scale as\cite{tsi03}: $1/T_{1}\sim s^5 (E_{\text{lh}}/\delta_0)^2 B^{3}$, where $s$ is the sound velocity and $E_{\text{lh}}/\delta_0$ is the size-dependent ratio of the heavy-light hole splitting and the isotropic exchange splitting (see Section \ref{sec:XMn}). For InAs/GaAs QDs at 15 T, this mechanism yields spin relaxation times of the order of a few nanoseconds, larger for smaller dots. For a CdTe dot, these times can be further decreased as a result of the sound velocity being smaller by about 40\% allowing to obtain values comparable to the exciton lifetime and in the same order of magnitude as measured experimentally.\cite{klo11pol} Moreover, the relaxation between the bright and dark states is expected to be even more efficient.\cite{ros07} Another mixing mechanism allowing for the phonon-assisted spin-flip is due to the spin-orbit coupling. This mechanism combined with piezo-electric spin-phonon coupling was shown to determine the relaxation rates of single electrons\cite{kha01,kro04} confined in QDs and to scale as $1/T_{1}\sim B^5$, providing an even more efficient process. Remarkably, in \CdMnTe\ QDs the {\em s-d} exchange interaction provides another important mechanism for mixing of the exciton states, namely via electron-Mn ion spin flip-flops (see Hamiltonian (\ref{Hsd}) and the discussion in Section \ref{sec:XMn}). Therefore, due to the combined action of the carrier-Mn exchange, electron-hole exchange, and the heavy-light hole mixing, the spin-up and spin-down exciton states are substantially mixed which in itself allows to expect short relaxation times. It also makes a rigid theoretical modeling challenging.

The second spin relaxation mechanism involves a dynamical role of the Mn ions. The {\em sp-d} exchange is no longer just a source of an effective quasi-static magnetic field giving rise to a splitting of the Kramers doublet, but may act as a scattering mechanism due to the dynamics of Mn spins caused by their short-range interactions (both isotropic and anisotropic superexchange). Such a role was invoked for spin relaxation of electrons in CdTe/\CdMnTe\ quantum wells.\cite{aki97} Proper description of the carrier (electron, hole, and exciton) spin relaxation in semimagnetic QDs obviously requires more theoretical investigations.

The model of the PL spectrum presented here is able to reproduce experimental data in a relatively wide range of external parameters such as magnetic field and temperature. The main effect not accounted for is the coexistence of different charge states in the single dot PL spectrum. As shown for CdTe QDs, even at small excitation powers all the $s$-shell transitions appear simultaneously in the spectrum, with a universal sequence $E(X^0)>E(X^+)>E(X^-)>E(2X)$, where $E(\chi)$ is the transition energy of the excitonic complex $\chi$.\cite{kaz11,klo11} For \CdMnTe\ dots with a sufficiently low Mn content it is possible to identify the neutral exciton and a charged exciton (see e.g., Fig. \ref{fig:tempe}), but due to the magnetization fluctuations, the broadening of the transitions result in charge exciton recombinations merged together with the biexciton transition. For dots with Mn concentration above 0.1, the transition linewidth ultimately precludes identification of the charge states. Our theoretical considerations assume a neutral QD and thus the PL spectrum model and discussion of the magnetic field dependence of the Zeeman shifts and linewidths concern exclusively the recombination of a neutral exciton. However, emission from a charged dot can exhibit a very different behavior in magnetic field, especially with respect to the development of carrier--Mn ion equilibrium. A positively charged dot, even with a very small Mn concentration, can reach the equilibrium due to the presence of the additional hole not taking part in the recombination and dwelling in the QD for a time much longer than both the exciton lifetime and EMP formation. This hole may develop a net magnetization due to the hole-Mn ion exchange interaction. For a dominant $X^+$ recombination we therefore expect the signatures of equilibrium such as the transition blueshift with increasing temperature or decreased Zeeman shift to appear even when the Mn density is low and EMP formation time is large. For a high Mn content, the picture becomes more complicated since both the single carrier and the exciton may develop the magnetization. However, while photocreation of an $X^0$ turns on the formation process, the excitation of the $X^+$ is rather expected to at least diminish the magnetization, since the two holes form a singlet states. Nevertheless, we expect to see different Zeeman shifts for a neutral and charged dot.

\section{Conclusions}

We have investigated the properties of the PL spectrum of semimagnetic, \CdMnTe\ QDs. We found that in the process of reaching a full thermal equilibrium in the exciton--Mn ion spin system, the PL spectrum undergoes changes in energy and linewidth. In particular, the equilibration process involves an intermediate stage, in which the exciton spin is relaxed, while the full exciton-Mn ion system is not yet in equilibrium. This particular stage gives rise to a specific shape of the spectrum, with a substantially narrowed and asymmetric transition line. We have developed a theoretical model, allowing to analytically compute the PL spectra for the unrelaxed, exciton spin-relaxed, and fully relaxed system. We found an excellent agreement between the model calculations and the experiment, and the comparison allowed to access such parameters as the exciton localization volume, the number of (approximately) paramagnetic Mn ions, and their temperature. The volumes evaluated from the experiment remain in agreement with AFM studies. The numbers of paramagnetic Mn ions in dots with different Mn densities agree well with those calculated assuming statistical distribution of the impurities in the cation sublattice. Furthermore, we have found that the heating of the Mn ions by the photocreated carriers is not effective for relatively large Mn densities although the Mn ions are still hotter than the bath. We have presented a comprehensive study of the PL energy and transition linewidth for \CdMnTe\ quantum dots in a broad range of Mn concentrations and discussed the processes controlling these parameters for various Mn densities.

This research was supported by Polish Ministry of Science and Higher Education grant no. 2011/01/B/ST3/022287 and by European Union within European Regional Development Fund, through Innovative Economy grant (POIG.01.01.02--00--008/08). \L.C. acknowledges support from the “FunDMS” Advanced Grant of the ERC within the “Ideas” 7th
Framework Programme. We thank Tomasz Dietl and Jacek Kossut for discussions.

\appendix

\section{Effective volume determined by interaction of the exciton with the Mn spins} \label{app:V}
Let us define the ratio of hole and electron volume as:
\begin{equation}
R = \frac{V_{h}}{V_{e}} \,\, ,
\end{equation}
and assume that $1/2 \! \leq \! R \! \leq \! 2$. We will now consider two cases: (i) Mn spins located only within the QD volume (which we identify here with the $V_{e}$) and (ii) Mn spins uniformly distributed in the QD and the surrounding barrier. Taking into account that in \CdMnTe\ $|\beta/\alpha|\! = \! 4$, the effective volume defined in Eq.~(\ref{eq:EV}) is then given by:
\begin{equation}
V = \frac{5 V_{h}}{4+R}
\label{eq:Va}
\end{equation}
in case (i) and $V\! = \! V_{h}$ in case (ii). Hence, for the assumed range of values for $R$, we get that in the former case $10/9 \cdot V_{h} \! \geq \!  V \! \geq \! 5/6 \cdot V_{h} \approx V_{h}$.

Note that there is an alternative definition of the effective volume, in which we take the Hamiltonian from Eq.~(\ref{eq:HXSz}) and use it to derive the expression for $\sigma_{E}$ analogous to the one from Eq.~(\ref{eq:sigma_E}):
\begin{equation}
\sigma_{E}^{2} = \frac{(\alpha-\beta)^2}{4V^{2}_{\sigma}}\sigma^{2}_{S} \,\, ,
\end{equation}
defining in this way the volume $V_{\sigma}$. A short calculation shows that in case (i) we have $V_{\sigma} \! = \! V$, while in case (ii) we have:
\begin{equation}
V_{\sigma} = \frac{5}{4}\frac{V_{h}}{\sqrt{\frac{3}{2}+\frac{R}{16}}} \,\, ,
\end{equation}
which gives $1.01 \cdot V_{\sigma} \! \geq \!  V \! \geq \! 0.98 \cdot V_{\sigma} \approx V_{h}$. This shows that both definitions of the effective volume are, with reasonable accuracy, consistent with each other and with the statement that the effective volume obtained from the fitting (Figs. (\ref{fig:mf-spectra},\ref{fig:mf})) corresponds to the confinement volume of the hole wavefunction.

\end{document}